\documentclass[referee]{aa}
\usepackage{psfig}

\usepackage{graphics}
\chardef\ii="10
\begin{document}
\thesaurus{08(09.10.1;09.13.2;08.16.5)}

\title{Intrinsic  colour calibration   
for F, G, K stars 
}
\author{Piercarlo Bonifacio\inst{1}
\and Elisabetta Caffau\inst{2}
\and Paolo Molaro\inst{1}
}
\offprints{P. Bonifacio}
\institute{
Osservatorio Astronomico di Trieste,Via G.B.Tiepolo 11, 
I-34131 Trieste, Italy
\and
Istituto Magistrale S.P.P. e L. annesso al Convitto Nazionale
Paolo Diacono, S. Pietro al Nat.
Udine, Italia
}
\mail{bonifaci@ts.astro.it}
\date{received .../Accepted...}
\titlerunning{Colour calibration}
\authorrunning{Bonifacio, Caffau \& Molaro}
\maketitle

\begin{abstract}
We derive an intrinsic colour calibration for F--K stars using broad
band Johnson colours and line indices $KP$ and $HP2$.
Through this calibration we can determine
$E(B-V)$ of an individual star 
within
0.03 mag.
The $E(B-V)$ values thus derived are in excellent agreement
with those derived from Str\"omgren photometry
through the Schuster \& Nissen (1989) calibration.
The agreement is also good with the reddening maps
of Burstein \& Heiles (1982) and Schlegel et al (1998), although
in this case there exists a small offset of about 0.01 mag.
This calibration may be applied to the large body
of data of the HK survey extension which will be
published in the near future.

\end{abstract}

\keywords{08.01.3 Stars: atmospheres - 08.07.1 Stars: general -
09.04.1 ISM:dust, extinction}
 
\section{Introduction}

An intrinsic colour calibration is a relation
defining, either in some colour-colour plane, or in some
N-dimensional space if the calibration requires N
colour indices, the locus  occupied by 
stars. The calibration is called
``intrinsic'' because it holds for the colours
which emerge from the stellar atmosphere but not
for the observed colours, which are altered while
passing through the interstellar medium. 
Knowledge of an intrinsic colour calibration
allows, at least in principle, to derive the reddening
of a given star by comparing  the observed colours
with the colours predicted by the calibration.
This may fail if the effect of reddening is to shift 
the point representative of a star along the 
curve, i.e. if the reddening vector is parallel
to the intrinsic locus. This occurs for instance
in the $\left( (B-V),(U-B)\right)$ plane for
F, G, K stars where the reddening
vector in the $\left( (B-V), (U-B)\right)$ plane is almost parallel
to the intrinsic locus of Main Sequence stars.

Determining the reddening of an individual
star is important for several reasons.
First, estimation of reddening
is necessary if we want
to obtain a photometric distance of the star.
Second, it is necessary for the derivation of atmospheric
parameters of the star, such as effective temperature,
surface gravity and metallicity. These parameters can be derived
from suitable colours, provided the reddening
is properly taken into account. For example
the $(V-K)$ colour may be accurately calibrated
onto $T_{eff}$ (Alonso et al  1996, 1999, Di Benedetto 1998),
but using $(V-K)$ rather than $(V-K)_0$ will result
in an underestimate of the temperature.
Finally the knowledge of the reddening gives
important information on the interstellar medium
along the line of sight towards the star.

For F, G and K stars the calibration of 
$uvby\beta$ colours by Schuster \& Nissen (1989)
has proved to be both a very powerful and accurate tool 
for estimating the reddening and has become the ``standard''
procedure for its determination.
The possibility of deriving such a calibration 
was foreseen in the very design of the
Str\"omgren photometric system and 
was already exploited by the calibrations
of Crawford (1975) and Olsen (1983) which
preceded the Schuster \& Nissen calibration and
are superseded by it.
The calibration is possible because the
system provides two indices, $(b-y)$ and $\beta$,
which mainly depend on effective temperature; 
however
while $(b-y)$ depends on reddening the
$\beta$ index is reddening--independent.
Therefore there exists a functional relation $(b-y)_0= f(\beta)$ which
allows to calibrate the reddening of the observed $(b-y)$.
This is accomplished by the
Schuster \& Nissen calibration in which terms in $m1$ allow to take
into account the metallicity dependence of $(b-y)$, while
terms in $c1$ allow to take into account its dependence
on surface gravity (luminosity).
Although very powerful, Str\"omgren photometry
requires a considerable investment in telescope time,
due to the large number of filters (6) 
and to their relatively narrow width.

The HK objective--prism/interference filter survey
(Beers et al 1985, 1992) provides, at present, the 
largest sample of stars suited for the study of the galactic
structure. The survey is kinematically unbiased and therefore
it is ideal for studying both kinematics and 
dynamics of the Galactic Halo.
Besides being the main source of extremely metal-poor stars,
[Fe/H]$< -3.0$, it provides a large number of stars
in the range $\rm -0.5 \le [Fe/H] \le -2.0$ which 
are well suited to study both the thick--disc and the 
halo thick--disc transition.
The medium dispersion follow--up survey, which provides
radial velocities and metallicities,
has been extended by Beers and collaborators
in both northern and southern hemispheres 
and the results will be soon available (see Beers
1999 for a summary). At the same time photometric campaigns
are being carried out to complement spectroscopic
data.  Norris et al (1999) provide $UBV$
data for $\sim 2500$ stars, Preston et al (1991) for about
1800 stars, Doinidis \& Beers (1990, 1991) for about 300 stars
and Bonifacio et al (2000) for about 300 stars.
Str\"omgren photometry is provided for 89 stars by
Schuster et al (1996), and for $\sim 500$ stars by
Anthony--Twarog et al (2000), although the latter data do
not include the $\beta$ index and  therefore cannot be used
to derive reddenings from the Schuster \& Nissen calibration.

From the above summary it is clear that the Schuster \& Nissen calibration
is of little use in determining reddenings for HK stars and would
require further observational efforts to obtain also
$uvby\beta$ data. So far reddenings for these stars
have been determined from maps, those of Burstein \& Heiles (1982)
in the first place, and, more recently, those of Schlegel et al (1998).
However it is possible to determine reddenings from available
data by developing a suitable Schuster \& Nissen -- type
calibration.
The indices involved in the Schuster \& Nissen calibration
are mostly measures of the following quantities:
slope of the Paschen continuum ($b-y$), metallicity ($m1$),
Balmer jump ($c1$), $H\beta$ ($\beta$).
Johnson photometry provides the slope of the Paschen continuum
($B-V$) and the Balmer jump ($U-B$), the line index 
$KP$  defined
in Beers et al (1999) is sensitive to  metallicity, while 
the index $HP2$ is a pseudo--equivalent--width of $H\delta$.  
It is therefore reasonable
to expect that a  Schuster \& Nissen -- type calibration,
involving $(B-V), (U-B), HP2$ and $KP$ may be derived.
In the following we show that this is indeed the case and that
reddening may be derived from it with an accuracy
comparable to that of the Schuster \& Nissen calibration.

\section{Derivation of the calibration}

To derive the calibration we 
used the sample of stars used by Beers et al (1999)
for the calibration of the $KP$
index. For all these stars
$UBV$ photometry and $HP2$ and $KP$ indices are available.
As calibrators we selected 
the stars  with low
reddening from  Beers et al (i.e. 
$E(B-V)\le 0.01$). 
To make this criterion more stringent and thus to select a sample
of truly unreddened stars,
we searched 
the Hauck \& Mermillod (1998) $uvby\beta$ catalog,
through the interface of
the General Catalog of
Photometric Data (GCPD, Mermillod, Hauck \& Mermillod, 1996) 
to  find out stars
of the  Beers et al sample with $uvby\beta$ photometry.
We then applied the Schuster \& Nissen calibration to these data to
obtain $E(b-y)\sim 0.74 E(B-V)$. Our final criterion was
that both the   estimate of $E(B-V)$ of Beers et al and
the one derived from Str\"omgren photometry were less than or equal to
0.01 mag\footnote{Since the Schuster \& Nissen 
calibration may result also
in negative values of $E(b-y)$ our criterion was $|1.35E(b-y)|\le 0.01$.}.

\begin{table}
\caption{Basic data for the calibrators}
\begin{center}
\renewcommand{\arraystretch}{0.6}

\begin{tabular}{llrrrrr}
\hline

STAR  & Type$^a$ & [Fe/H]    &  $(b-y)$ & $ m1$    &  $c1$    & $ \beta$   \\
\hline
\\
 BD +18\degr 1479 & D  &  --0.46 &  0.440 &  0.258 &  0.256 &  2.560  \\ 
 CD --24\degr 17504$^b$ & D &  --3.70 &  0.317 &  0.039 &  0.301 &  2.596  \\ 
 G8--16        & D &  --1.59 &  0.322 &  0.071 &  0.292 &  2.595  \\ 
 G11--36       & D & --0.68 &  0.384 &  0.139 &  0.255 &  2.579 \\ 
 G11--44       & D & --2.07 &  0.334 &  0.054 &  0.263 &  2.597  \\ 
 G11--45       & D & --0.01 &  0.431 &  0.249 &  0.359 &  2.602 \\ 
 G12--21       & D & --1.32 &  0.339 &  0.090 &  0.284 &  2.592  \\ 
 G13--35       & D & --1.63 &  0.331 &  0.060 &  0.285 &  2.594 \\ 
 G14--23       & D & --0.27 &  0.370 &  0.150 &  0.316 &  2.590  \\ 
 G14--26       & D & --0.20 &  0.387 &  0.181 &  0.349 &  2.590  \\ 
 G15--6        & D & --0.65 &  0.436 &  0.210 &  0.253 &  2.561 \\ 
 G15--17       & D & --0.39 &  0.475 &  0.315 &  0.246 &  2.553 \\ 
 G17--21       & D & --0.66 &  0.367 &  0.125 &  0.306 &  2.588  \\ 
 G17--30       & D & --0.48 &  0.396 &  0.164 &  0.303 &  2.574  \\ 
 G24--15       & D & --1.10 &  0.342 &  0.092 &  0.273 &  2.597  \\ 
 G37--26       & D & --1.93 &  0.351 &  0.058 &  0.208 &  2.584 \\ 
 G44--6        & D & --0.54 &  0.397 &  0.160 &  0.259 &  2.566  \\ 
 G44--30       & D & --0.89 &  0.427 &  0.177 &  0.200 &  2.557 \\ 
 G54--21       & D & --0.03 &  0.387 &  0.195 &  0.319 &  2.602  \\ 
 G57--11       & D &  0.03 &  0.412 &  0.230 &  0.328 &  2.593  \\ 
 G58--23       & D & --0.97 &  0.407 &  0.138 &  0.219 &  2.565 \\ 
 G58--25       & D & --1.41 &  0.344 &  0.079 &  0.258 &  2.589  \\ 
 G58--41       & D & --0.33 &  0.371 &  0.150 &  0.344 &  2.591  \\ 
 G59--1        & D & --1.02 &  0.424 &  0.186 &  0.204 &  2.567 \\ 
 G59--24       & D & --2.42 &  0.332 &  0.054 &  0.225 &  2.593  \\ 
 G60--48       & D & --1.63 &  0.365 &  0.074 &  0.189 &  2.583  \\ 
 G62--44       & D & --0.58 &  0.473 &  0.295 &  0.248 &  2.554 \\ 
 G62--52       & D & --1.28 &  0.430 &  0.167 &  0.190 &  2.553 \\ 
 G63--46       & D & --0.91 &  0.382 &  0.139 &  0.283 &  2.584 \\ 
 G65--47       & D & --0.35 &  0.405 &  0.188 &  0.281 &  2.584 \\ 
 G66--15       & D & --0.20 &  0.405 &  0.204 &  0.319 &  2.582  \\ 
 G80--15       & D & --0.78 &  0.365 &  0.126 &  0.272 &  2.586  \\ 
 G88--32       & D & --2.36 &  0.309 &  0.051 &  0.357 &  2.591  \\ 
 G97--43       & D & --0.49 &  0.463 &  0.256 &  0.277 &  2.554  \\ 
 G113--24      & D & --0.49 &  0.383 &  0.146 &  0.301 &  2.588 \\ 
 G114--26      & D & --1.78 &  0.349 &  0.076 &  0.250 &  2.594 \\ 
 G119--64      & D & --1.42 &  0.319 &  0.073 &  0.319 &  2.600  \\ 
 G121--12      & D & --0.92 &  0.350 &  0.094 &  0.287 &  2.592  \\ 
 G160--3       & D & --0.14 &  0.414 &  0.217 &  0.339 &  2.584  \\ 
 G162--16      & D & --0.53 &  0.396 &  0.167 &  0.323 &  2.586  \\ 
 G162--51      & D & --0.52 &  0.377 &  0.145 &  0.305 &  2.583  \\ 
 G162--68      & D & --0.54 &  0.425 &  0.202 &  0.206 &  2.565  \\ 
 G165--39      & D & --2.05 &  0.309 &  0.055 &  0.354 &  2.602 \\ 
 G200--62      & D & --0.45 &  0.479 &  0.304 &  0.273 &  2.552 \\ 
 G229--34      & D & --0.50 &  0.403 &  0.190 &  0.334 &  2.579  \\ 
 G271--34      & D & --0.68 &  0.386 &  0.155 &  0.262 &  2.576  \\ 
 HD 693       & TO & --0.38 &  0.328 &  0.130 &  0.405 &  2.621  \\ 
 HD 3567      & SG & --1.29 &  0.328 &  0.089 &  0.330 &  2.600  \\ 
 HD 6461      & G  & --0.93 &  0.500 &  0.199 &  0.434 &  2.554  \\ 
 HD 16031     & D  & --1.71 &  0.323 &  0.069 &  0.302 &  2.606 \\ 
 HD 20010     & SG & --0.27 &  0.339 &  0.156 &  0.411 &  2.624  \\ 
 HD 34328     & D  & --1.61 &  0.365 &  0.063 &  0.205 &  2.569  \\ 
 HD 76932     & D  & --0.99 &  0.359 &  0.119 &  0.298 &  2.584  \\ 
 HD 90508     & D  & --0.23 &  0.397 &  0.175 &  0.298 &  2.578  \\ 
 HD 105590    & HB & --0.17 &  0.414 &  0.229 &  0.321 &  2.591  \\ 
 HD 113083    & D  & --1.04 &  0.367 &  0.122 &  0.254 &  2.587  \\ 
 HD 114762    & D  & --0.70 &  0.365 &  0.125 &  0.297 &  2.588 \\ 
 HD 134169    & SG & --0.85 &  0.370 &  0.119 &  0.312 &  2.582  \\ 
 HD 184499    & D  & --0.58 &  0.390 &  0.143 &  0.314 &  2.578  \\ 
 HD 193901    & D  & --1.08 &  0.381 &  0.103 &  0.217 &  2.573  \\ 
 HD 200580    & D  & --0.75 &  0.364 &  0.149 &  0.305 &  2.599  \\ 
 HD 201889    & D  & --0.92 &  0.388 &  0.148 &  0.281 &  2.577  \\ 
 HD 201891    & D  & --1.13 &  0.358 &  0.104 &  0.262 &  2.590 \\ 
 HD 205156    & D  & --0.57 &  0.398 &  0.161 &  0.256 &  2.573  \\ 
 HD 219617    & D  & --1.31 &  0.344 &  0.078 &  0.246 &  2.597 \\ 
\\
\hline
\end{tabular}
\end{center}
\noindent $^a$ D=dwarf; TO = turn--off; SG=subgiant; G=giant; 
HB=horizontal branch.

\noindent $^b$ not used in the calibration, because trimmed out after the first
pass.
\end{table}

Out of the sample of Beers et al (1999) 65 stars satisfy our criterion
and they are reported in Table 1, the star name is given in column (1),
column (2) gives the star type, according to Beers et al (1999).
Column (3) is [Fe/H] and columns (4)--(7) provide the Str\"omgren
photometry extracted from the Hauck \& Mermillod (1998) catalogue.
We performed a $\chi^2$ fit
on this sample of stars for several functional forms. We 
computed the rms of the fit and we then discarded 
those stars whose residual was greater than $2.5\times$ rms.
This was aimed at further cleaning the sample by rejecting 
stars which are either
reddened or whose colours or line indices are affected by 
larger errors. Quite interesting only one star, 
CD --24\degr 17504, was discarded,
whichever functional form was used. 
This star is also the most metal--poor of the sample, in fact
the only one below [Fe/H]=--3.00.
Thus our fits were all
performed on a sample totalling 64 stars.

We began with the assumption that both ($B-V$) and $HP2$
strongly depend on temperature; we therefore fit a linear relation

$$ (B-V) = x_1 + x_2 \log HP2$$ 

and obtained a  fit with $\chi^2 = 273.8$. We next began to
add other terms in $\log KP$ and ($U-B$) and checked the significance
of the new term through the F test (Bevington \& Robinson, 1992, p. 208).
Linear terms in both $\log KP$ and ($U-B$)  are highly significant
(the probability  corresponding to the observed $\rm F_\chi $
is 
$6.5\times 10^{-7}$ for $\log KP$ and $7.0 \times 10^{-13}$ for 
($U-B$)). On the other hand further quadratic terms are not significant,
the most significant being a term in $(\log KP)^2$, with a 
probability of the observed $\rm F_\chi \approx 5.5 \times 10 ^{-2}$, i.e.
a significance
of $\sim 95 \%$. We therefore conclude   that the best functional 
form is that with only linear terms.

\begin{equation}
\label{eqfit}
(B-V)_0 = x_1 + x_2 \log HP2 +x_3 \log KP + x_4 (U-B)_0 
\end{equation}

Our best fit parameters are given in Table 2, together with their 
formal errors 
derived from the covariance matrix.

\begin{table}
\caption{Best fit parameters}
\begin{center}
\begin{tabular}{rrr}
\hline
 & param. & error \\
\hline

$x_1$ &$ 0.5734$ &0.0004\\
$x_2$ &$-0.2759$ &0.0006\\
$x_3$ &$ 0.1040$ &0.0003\\
$x_4$ &$ 0.2676$ &0.0006\\
\hline
\end{tabular}
\end{center}
\end{table}

\begin{figure*}
\psfig{figure=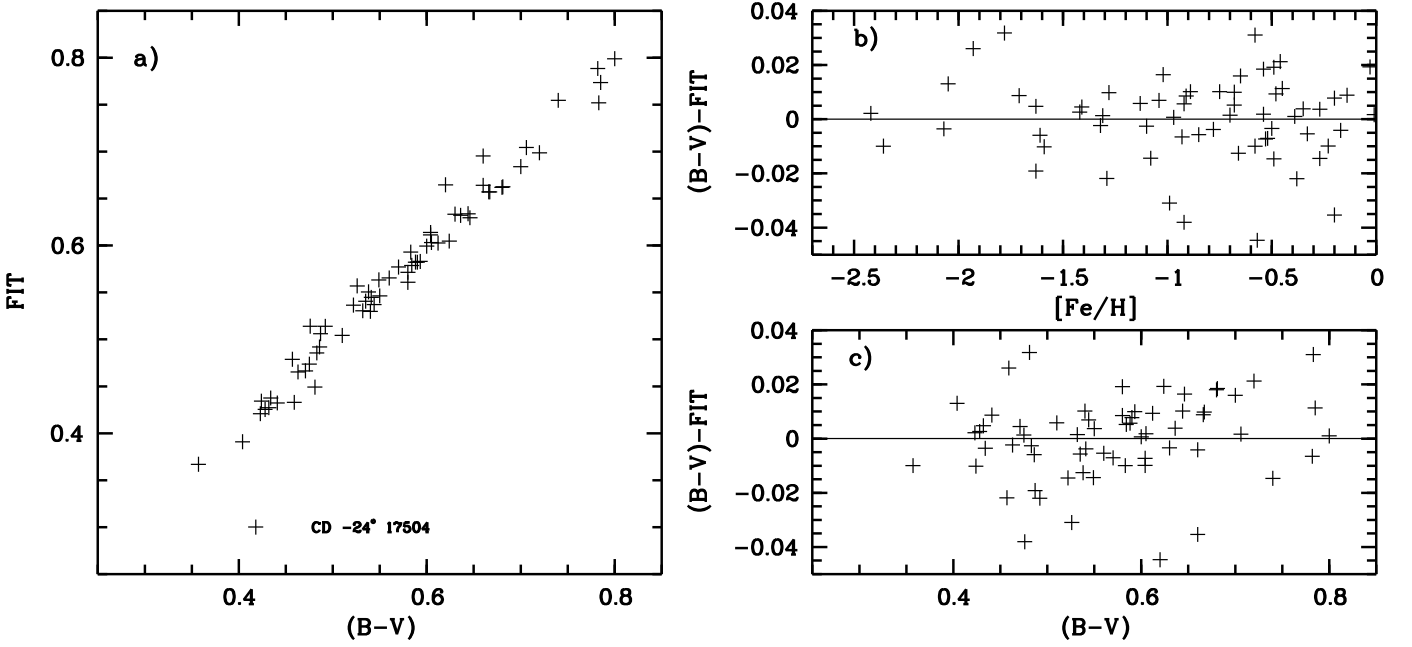,clip=t}
\caption{
{\bf a)} The right hand side of equation \ref{eqfit} for
the calibrators as a function of ($B-V$).
The star CD --24\degr 17504 appears to be an outlier and
has not been used in the derivation of the calibration, and it is not
drawn in panels {\bf b)} and {\bf c)}.
{\bf b)} The residuals ($B-V$)--FIT as a function of metallicity.
{\bf c)} Same as panel {\bf b)} but as a function of
($B-V$).
}
\label{fit}
\end{figure*}

The rms of the fit was 0.0153 mag and $\chi^2 \approx 67.7$,
i.e. $\chi^2_{60}= 1.13$, which indicates a good fit.
In Fig. \ref{fit}, panel a) we show a plot of $(B-V)$
versus the right hand side of Eq. \ref{eqfit}. 
Panels b) and c) display the residuals as a function
of metallicity and $(B-V)$ colour, respectively.

The range of validity of the calibration is
fixed by the properties of the calibrator stars. In our case
these are:

$$
\matrix{
\phantom{-}0.375 &\le& (B-V)\hfill& \le& 0.800\cr
\phantom{-}0.780 &\le& HP2\hfill & \le & 4.330\cr
\phantom{-}0.186 &\le&  KP\hfill & \le & 9.790\cr
-0.245&\le& (U-B)\hfill &  \le& 0.350
}
$$

The mean metallicity of the calibrators is [Fe/H] $\sim -0.9$.
The metallicity range is $\rm -2.42\le [Fe/H] \le +0.03$.
In practice we expect that our calibration is applicable
to stars of intermediate metallicity. 
We note that the mean metallicity of the calibrators of 
Schuster \& Nissen is [Fe/H]$=-0.50$, 
while their metallicity range is $\rm -2.49\le [Fe/H] \le +0.22$.
Thus there is an almost perfect overlap of the metallicity domains 
where
the two calibrations are derived, with the Schuster \& Nissen
calibration extending towards slightly higher metallicities.

A matter of concern is wether the luminosity (gravity) dependence
of the calibration is properly captured by the $(U-B)$ colour.
Inspection of Table 1 reveals that, although we did not impose
any selection criterion on luminosity,  most of our stars are dwarfs, 
only one is a giant, three sub--giants and one horizontal branch.
This is a result of imposing a very tight criterion on reddening
for the calibrators:
all low--reddening stars are nearby and therefore most are dwarfs.
So formally our calibration is valid only for dwarfs. 
In Fig. \ref{c1by} ~ we show the residuals of the fit
as a box plot in the $\left((b-y)_0,c0\right)$ plane.   
There appears to be no obvious trend of the residuals with the
luminosity of the stars, we may therefore expect that 
the calibration is in fact equally applicable to dwarfs and
giants, as is the Schuster \& Nissen calibration.

\section{Reddening estimation and comparison with other estimates}

Equation \ref{eqfit} can be used to determine $E(B-V)$ once all
the observed indices are known. Since it is a first order equation both in 
$(B-V)$ and $(U-B)$ we may solve analytically for $E(B-V)$,
rather than by iteration, as it is necessary with the Schuster \&
Nissen calibration. By adopting the reddening slope
$E(U-B)/E(B-V)=0.72$
we obtain:

\begin{equation}
\label{ebv}
E(B-V) = {(B-V)-x_1-x_2\log HP2 -x_3\log KP - x_4 (U-B)\over (1-0.72x_4)}
\end{equation}

We can now use Eq. \ref{ebv} to compute $E(B-V)$ for stars
for which independent estimates of reddening are available, in order
to asses the external error in the derived reddening. 
The internal error is of the order
of the rms of the calibration, 0.015 mag.

\begin{figure}
\psfig{figure=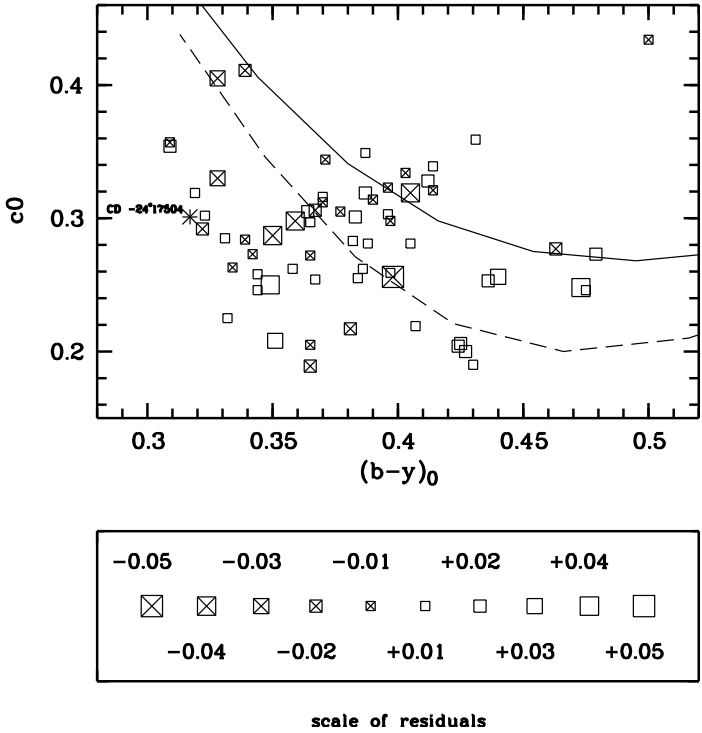,clip=t}
\caption{
The 
residuals ($B-V$)--FIT in the $\left( (b-y)_0,c0\right)$ plane.
The stars have been divided into 10 bins
0.01 mag wide.
The size of the symbol is largest for the stars 
in the bins $0.04< |(B-V)-{\rm FIT}|\le 0.05$ mag and decreases
to the minimum size for stars in the bins
with $|(B-V)-{\rm FIT}|\le 0.01$,
as indicated in the scale plot  shown below the figure. 
Negative residuals are shown 
with crossed squares, while positive residuals are shown
with open squares. The star CD --24\degr 17504  
is drawn with an asterisk.
The solid line represents the locus of points with $\log g=3.5$
for [Fe/H]=--1.0. The dashed line is the same but for 
[Fe/H]=--2.5.
} 
\label{c1by}
\end{figure}

\begin{figure*}
\psfig{figure=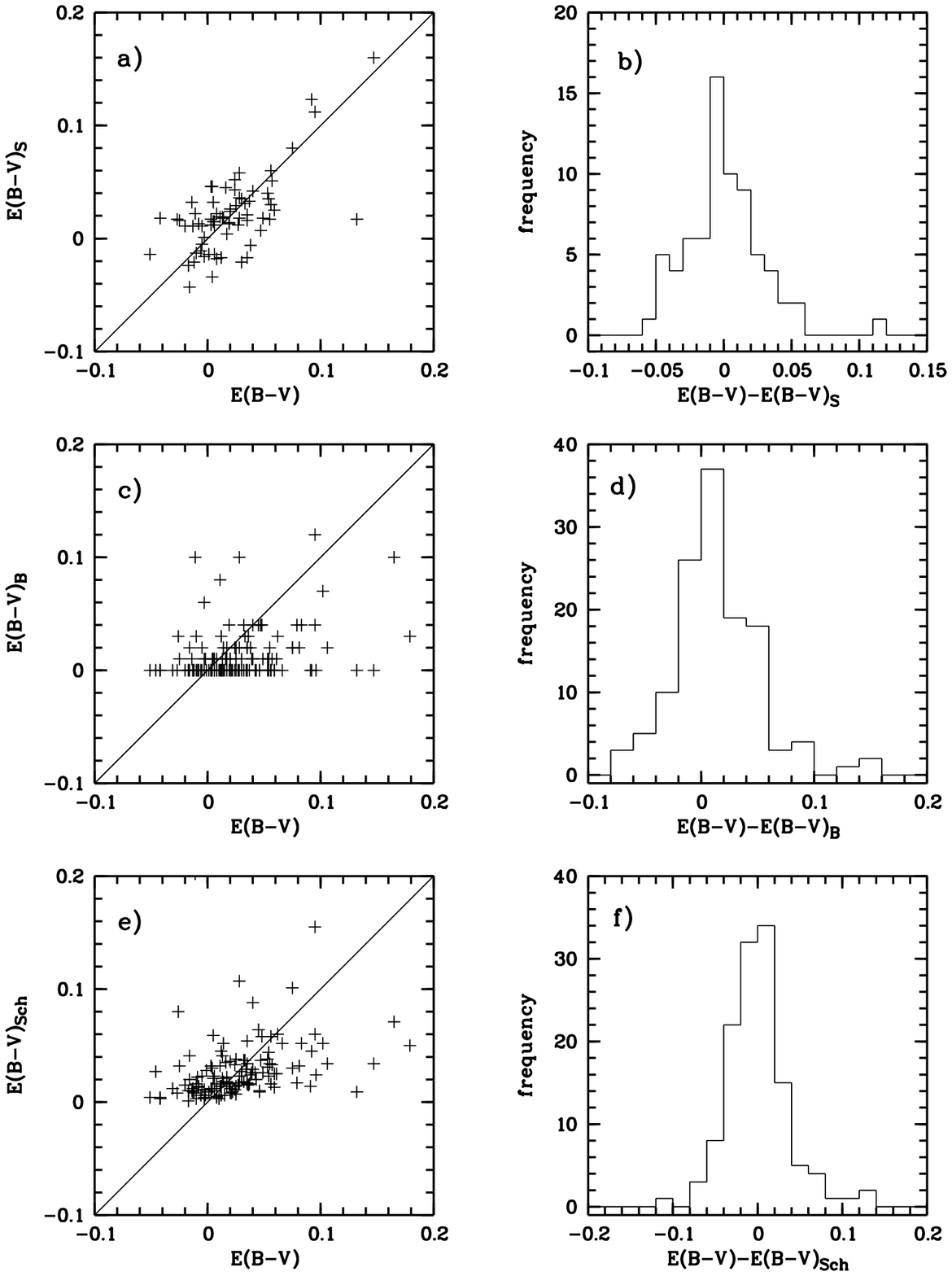,clip=t}
\caption{Comparison of the reddening derived from equation \ref{ebv}
with that derived from Str\"omgren photometry (panels a) and b)),
with that of Beers et al (panels c) and d))
and with that derived from the maps of Schlegel et al
(panels e) and f)).}
\label{compb}
\end{figure*}

The Beers et al (1999) sample may be conveniently used for comparison.
From the sample we exclude all the calibrators and keep
all the stars which had been rejected because either their reddening
was too large
or they lacked Str\"omgren photometry. Out of this sample
we further selected only the stars with indices within the range 
of the calibration.  The sample of comparison stars now consists
of 129 stars, of which 71 have also Str\"omgren photometry.
We  start by comparing the reddening derived from Eq. \ref{ebv}
with that derived from the Schuster \& Nissen calibration
through $E(B-V)=1.35E(b-y)$. The result of the comparison is
shown in panel a) of Fig. \ref{compb}.

\begin{figure}
\psfig{figure=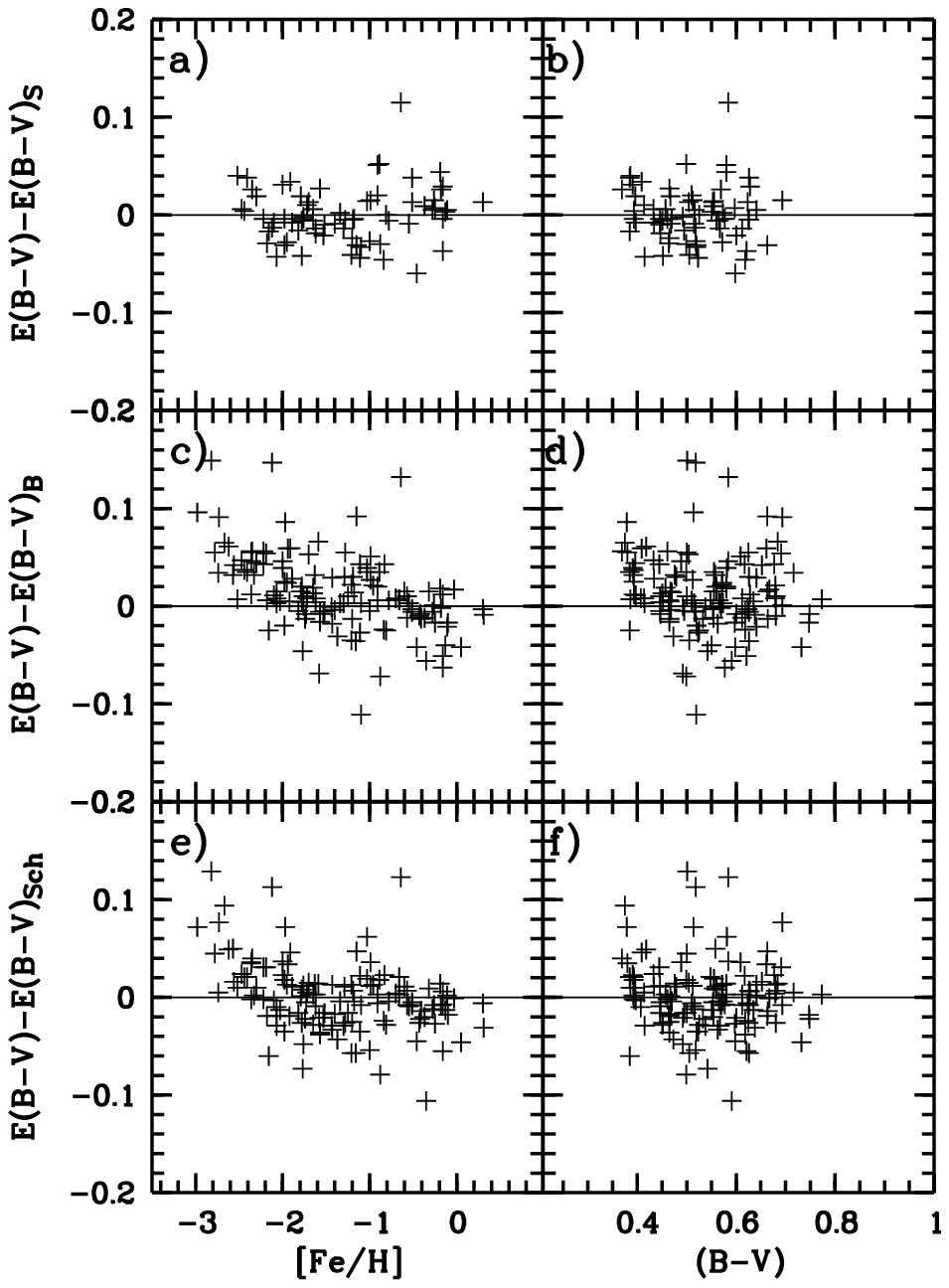,clip=t}
\caption{Differences $E(B-V)_{(our)}-E(B-V)_{(other)}$ 
as a function of metallicity and $(B-V)$ colour.
In panels a) and b) our reddening is compared to that derived
from the Str\"omgren photometry; in panels c) and d) to that
given by Beers et al; in panels e) and f) to that
derived form the maps of Schlegel et al.
}
\label{resid}
\end{figure}

A clear outlier may be noticed (HD 7424), for which our reddening estimate
is more than 0.1 mag larger than that derived from Str\"omgren photometry.
By dropping this star our sample eventually includes
70 comparison stars.
The histogram of the difference ($E(B-V)_{ours}- E(B-V)_{S}$)
is shown in panel b) of Fig. \ref{compb}. 
In Fig. \ref{resid}, panels a) and b), 
we display the differences as a function
of [Fe/H] and $(B-V)$,  no trend with either is apparent.
The mean value of the difference is almost
zero ($0.002$ mag) and the standard deviation is 0.025 mag.
We note that if HD 7424 is kept in the sample only
a slightly larger standard deviation of 0.028 mag would result.
This shows that the reddening derived from our calibration is in
good agreement with that derived from the Schuster \& Nissen calibration.
The value of 0.025 may be regarded as an error estimate
of the reddening derived through equation \ref{ebv}.

Next we compare our reddening with that reported by Beers et al(1999),
which is mostly based on the Burstein \& Heiles (1982) reddening
maps (see Beers et al. 1999 for further details on their adopted
reddening).
The plot in which the reddenings are compared and the histogram of
the differences ($E(B-V)_{ours} - E(B-V)_{B}$)
are shown in panels c) and d) of Fig. \ref{compb}, respectively.
There is an evident offset between the two reddening
estimates  as well as a tail with large differences.
This is made up of three stars: HD 7424, already
identified as an outlier in the comparison with reddening from
Str\"omgren photometry; HD 161770, for which Beers et al (1999) 
give a zero reddening while
we obtain 0.147 from equation \ref{ebv} and 0.160 from
the  Schuster \& Nissen calibration; G82-23 for which 
Beers et al (1999) give  0.03
while we obtain
0.179, where no Str\"omgren
photometry is available for this star.
The mean difference for the whole sample (129 stars) 
is 0.014 mag and the standard deviation is 0.040 mag.
If we remove the three above-mentioned stars from the sample 
the mean difference becomes 0.011 mag and the standard deviation
0.034 mag. 
In Fig. \ref{resid} panel c) we may notice
a slight trend of the differences
with metallicities, while no such trend appears with $(B-V)$ colour.

Finally, we compare our reddenings with those derived from the 
recent reddening maps of Schlegel et al (1998). 
In order to obtain the 30\%--50\% reddening reduction recommended by
Arce \& Goodman (1999) for the highly reddened stars
we  modify the reddenings
above 0.10 mag as described in Bonifacio, Monai \& Beers (2000)
\footnote{For stars for which
the reddening predicted by the Schlegel et al (1998) maps  
$E(B-V)_{Sch}> 0.1$ 
Bonifacio et al (2000) adopt $E(B-V) = 0.10 +0.65\times[E(B-V)_{Sch} -0.1]$.}.
Several of our comparison stars are quite close and therefore within
the dust layer.  
The reddening provided by the maps, refers instead to the full
line of sight and should be applied as it stands only to extragalactic
objects or to objects well above the dust layer.
We take this into account by multiplying the reddening
of the maps by a factor $[1-\exp(-|d\sin b|/h)]$, where $d$ is the star's
distance $b$ its galactic latitude and $h$ the scale height of the
dust layer, which we assumed to be 125 pc.
The distances were taken from Beers et al (1999).
The plot of the reddening obtained from the Schlegel et al (1998) maps
versus our reddening is shown in panel e) of Fig. \ref{compb} and
the histogram of the differences in panel f). 
The mean value of the difference ($E(B-V)_{our} - E(B-V)_{Sch}$)
is 0.009 mag and the standard deviation is 0.041 mag.
Five stars out of the sample have absolute difference 
larger than 0.1 mag, namely G82-23, HD 7424 and
HD 161770 have a difference $>0.1$, while G79-42 and G99-40
have a difference $<-0.1$. The reddening predicted by the Schlegel  et
al maps for the latter two stars is very high, in spite of our 
reduction (0.411 for G79-42 and 0.707 for G99-40). If we treat these
five stars as outliers and recompute both the mean and the
standard deviation we obtain 0.013 mag and 0.030 mag respectively.
In Fig. \ref{resid} panel e), shows a slight dependence 
of the differences on metallicities, while panel f) shows no
trend with $(B-V)$.

\begin{figure}
\psfig{figure=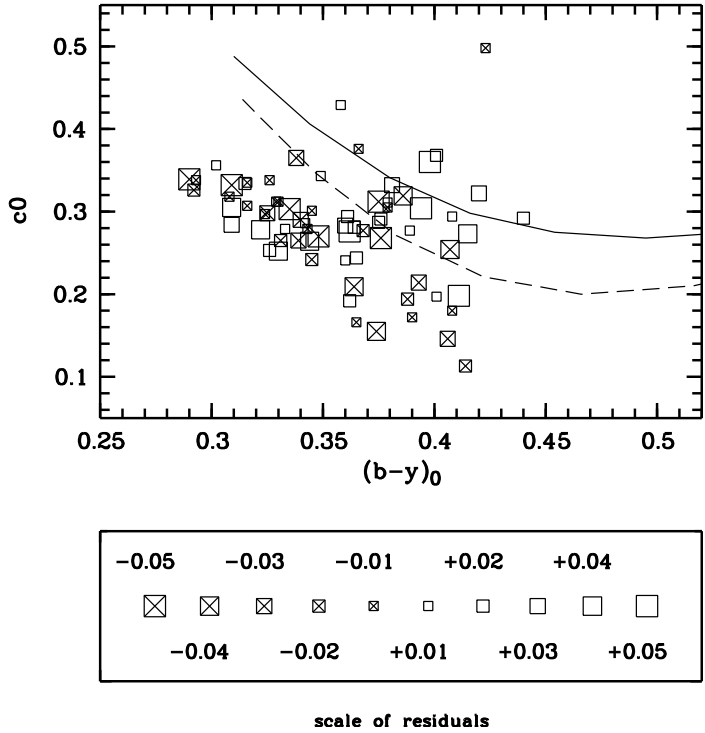,clip=t}
\caption{
The 
residuals $E(B-V)$--$E(B-V)_S$ in the $\left( (b-y)_0,c0\right)$ plane
for the comparison stars.
The stars have been divided into 10 bins
8 are 0.01 mag wide,  one is $E(B-V)-E(B-V)_S<-0.04$
and one $E(B-V))-E(B-V)_S>+0.04$. The rest of
the symbols are as in Fig \ref{c1by}.
} 
\label{c1byconf}
\end{figure}

Our final check is on the possible dependence of the calibration
on the luminosity of the stars. In Fig \ref{c1byconf} we show 
a box plot of the residuals in the $\left((b-y)_0,c0\right)$ plane,
similar to Fig \ref{c1by}, but for the comparison stars, rather
than for the calibrators. Among the comparison stars with
Str\"omgren photometry there is only one giant 
seven subgiants and two horizontal branch stars, the rest are
dwarfs or turn--off stars. Although the high luminosity stars
are under--represented there does not appear to be any  
obvious trend in the residuals with gravity.

From the above discussion we conclude that our reddening is
comparable to  that derived from Str\"omgren photometry
through the Schuster \& Nissen calibration, while with respect
to the reddening derived from the maps (either those of Schlegel et al
or those of Burstein \& Heiles) there is an offset of $\sim 0.01$ mag,
in the sense that the reddening predicted by equation \ref{ebv}
is higher than that predicted by the maps.
The comparison also suggests that the accuracy of our reddening
estimate is of the order of 0.03 mag, and therefore it is comparable to 
the reddening  obtained from the Schuster \& Nissen calibration.

\section{Conclusions}

We derived an intrinsic colour calibration involving
the broad band colours $(B-V)$ and $(U-B)$ and the line indices
$HP2$ and $KP$. Within its validity range the calibration may be used
to derive the reddening $E(B-V)$ of individual stars
with an accuracy of the order of 0.03 mag.
We showed that the reddenings derived from our calibration
are in good agreement with those derived from the Schuster \& Nissen (1989)
calibration of Str\"omgren photometry.
When compared to  reddening determinations based on maps, either
those of Burstein \& Heiles (1982) or those of Schlegel et al (1998),
our reddenings are systematically higher by about 0.01 mag.

The data base for $HP2$, $KP$, $(B-V)$  and $(U-B)$
for HK survey stars is steadily increasing thanks to the efforts
of Beers and collaborators and presently data for about 3000 stars
are available. Our calibration may be used to derive
reddenings for all these stars. Dereddened colours are necessary
to determine the metallicity from the Beers et al (1999) calibration
and to derive a photometric distance.
In the case of stars which will be observed spectroscopically at high
resolution, dereddened colours will allow the determination of effective
temperatures from colour--temperature calibrations.

Whenever $uvby\beta$ photometry is available the Schuster \& Nissen
calibration is the preferred way to determine the reddening of a star.
However, when no Str\"omgren photometry is available, our calibration
provides a useful way to estimate reddening with comparable accuracy.

\begin{acknowledgements}
We are grateful to the referee, Dr. A. Alonso, for carefully
reviewing   our paper  and for suggesting many additions and improvements
to its first version.
This research made use of the SIMBAD data base operated
at CDS, Strasbourg.
\end{acknowledgements}

\end{document}